\documentclass{iau} 
\usepackage{units}
\usepackage{natbib}
\usepackage{graphicx}

\usepackage{apjfonts,graphicx,graphics}

\usepackage{hyperref}
\usepackage{mathptmx}
\usepackage{amsmath}
\usepackage{wasysym}
\usepackage{times}
\usepackage{epsfig}
\usepackage{ulem}
\usepackage{epstopdf}
\usepackage{appendix}

\usepackage{import}
\newbox{\bigpicturebox}
\usepackage[capitalize]{cleveref}
\usepackage{CJKutf8}

\title[Shock wave signature in non-fundamental mode RR Lyrae] 
{Detecting shock waves in non-fundamental mode RR Lyrae using large sample of spectra in SDSS and LAMOST}

\author[Xiao-Wei Duan \textit{et al.}]   
{Xiao-Wei Duan$^{1,2,3}$, Xiao-Dian Chen$^{3,4,6}$, Li-Cai Deng$^{3,4,1,6}$, Fan Yang$^{3}$, Chao Liu$^{4,5}$, Anupam Bhardwaj$^{2}$, Hua-Wei Zhang$^{1,2}$}

\affiliation{$^1$Department of Astronomy, Peking University, Yi He Yuan Road 5, Hai Dian District, Beijing 100871, China \\ email: {\tt duanxw@pku.edu.cn} \\[\affilskip]
$^2$Kavli Institute for Astronomy \& Astrophysics, Peking University, Yi He Yuan Road 5, Hai Dian District, Beijing 100871, China \\[\affilskip]
$^3$CAS Key Laboratory of Optical Astronomy, National Astronomical Observatories, Chinese Academy of Sciences, Beijing 100101, China \\[\affilskip]
$^4$School of Astronomy and Space Science, University of the Chinese Academy of Sciences, Huairou 101408, China \\[\affilskip]
$^5$CAS Key Laboratory of Space Astronomy and Technology, National Astronomical Observatories, Chinese Academy of Sciences, Beijing 100101, China \\[\affilskip]
$^6$Department of Astronomy, China West Normal University, Nanchong 637009, China \\[\affilskip]
}

\begin{document}

\maketitle

\begin{abstract}
Steps toward the nature inside RR Lyrae variables can not only improve our understanding of variable stars but also innovate the precision when we use them as tracers to map the structure of the universe. In this work, we develop a hand-crafted one-dimensional pattern recognition pipeline to fetch out the ``first apparitions'', the most prominent observational characteristic of shock. We report the first detection of hydrogen emission lines in the first-overtone and multi-mode RR Lyrae variables. We find that there is an anti-correlation between the intensity and the radial velocity of the emission signal, which is possibly caused by opacity changing in the helium ionization zone. Moreover, we find one RRd star with hydrogen emission that possibly shows Blazhko-type modulations. According to our discoveries, with an enormous volume of upcoming data releases of variable stars and spectra, it may become possible to build up the bridge between shock waves and big problems like the Blazhko effect in non-fundamental mode RR Lyrae stars.
 
\end{abstract}

\firstsection 
\section{Introduction \label{sec:intro}}
RR Lyrae stars represent the low-mass (0.5-0.7 $M_{\odot}$), old-age (>10 Gyr) stellar population at the core helium (He)-burning stage of their evolution. Their pulsations are caused by $\kappa$-mechanism when the opacity of ionized He changes with the temperature. They obey the absolute visual magnitude versus metallicity law $M_{V}-[Fe/H]$$ $ \citep{Muraveva2018MNRAS.481.1195M} and the period-luminosity-metallicity law \citep[PLZ,][]{Longmore1986MNRAS.220..279L,Catelan2004ApJS..154..633C} in the infrared bands, thus making them excellent standard candles for distance determinations to nearby galaxies. RR Lyrae stars are further classified based on the number of oscillation modes, as fundamental mode (RRab), the first overtone mode (RRc), and multi-mode (RRd) variables \citep{Soszy2011AcA....61....1S}. Among them, RRc and RRd stars enjoy shorter periods and smaller amplitudes than RRab and occupy the blue side of RR Lyrae instability strip.

It is now a well-established fact that for RRab stars, three moderate-to-small emission lines appear sequentially in a pulsation cycle, which are the so-called ``three apparitions'' \citep{Preston2011AJ....141....6P}. They are considered to be generated by atoms de-exciting after being excited by the shock wave, which is compressing, heating, and accelerating the atmospheric gas when traveling into it \citep{Gillet2014A&A...565A..73G}. The ``first apparitions'' are generated when the shock wave is close to the photosphere, which indicates the coming of maximum luminosity. The ``second apparitions'' are thought to be produced by the photospheric compression, when the inner deep expanding atmosphere collides with outer layers which are in ballistic infalling motion \citep{Hill1972}. 
\cite{Chadid2013MNRAS.434..552C} interpreted the ``third apparitions'' as to be generated by the shock wave $Sh_{PM}$, a superposition of the compression resulted from the hydrogen recombination front, and an accumulation of several weak compression waves. \cite{Gillet2019} provided a general overview of the atmospheric dynamical structure of RRab over a typical pulsating cycle, using high-resolution spectra of $H\alpha$ and sodium lines in RR Lyr (HD 182989), which describes the evolution picture of shock waves inside quite clearly.

Although the ``apparitions'' were generally detected in RRab, there is no detection of shock waves in RRc or RRd stars. \cite{Gillet2013A&A...554A..46G} suggested that the intensity of the shock waves is certainly lower in RRc than RRab. And it is possible that the coexistence of stable oscillations in fundamental mode and the first-overtone mode in the same pulsation cycle can reduce the development of the shock amplitude. Up to now, there hasn't been any atmospheric pulsation model for RRc or RRd with a comprehensive study of the dynamical evolution of shock waves. The research about the link between the Blazhko effect and the interaction of strong shock waves was strongly obstructed by this absence.  

In this work, we develop a large sample searching algorithm to investigate shock waves in RR Lyrae stars, especially those in non-fundamental mode. We focus on the ``first apparition'', which is the most prominent signature of shock waves to be observed and allows for large database searching. They show up as emissions on the blue wings of Balmer lines when $\phi\sim0.9$. We select this kind of features through our hand-crafted one-D pattern recognition pipeline, using low-resolution and single-epoch spectra. Here we report the first detection of hydrogen emission in the first-overtone and multi-mode RR Lyrae stars, which give us new insights into the nature of non-fundamental mode RR Lyrae variables.

\section{Observations and methods}

We need photometric and spectroscopic observations. Photometry was used to classify RR Lyrae stars, which can be provided by the Catalina Sky Surveys \citep{Drake2014,Drake2017}, the Wide-field Infrared Survey Explorer \citep[WISE,][] {chen2018}, the All Sky Automated Survey for Supernovae \citep[ASAS-SN,][]{Jayasinghe2019MNRAS.486.1907J}, and the Asteroid Terrestrial-impact Last Alert System \citep[ATLAS,][]{Heinze2018AJ....156..241H}. We combine these catalogs, remove duplicate data, and get 68,152 stars, which are identified as RR Lyrae. Light curves from the Zwicky Transient Facility \citep[ZTF,][]{Bellm2019PASP..131a8002B,Chen2020} for our selected stars are collected if available.

We search for the pattern of the ``first apparitions'' among low-resolution and single-epoch spectra. The spectra are collected from the Sloan Digital Sky Survey \citep[SDSS,][]{Eisenstein2011AJ....142...72E} and the Large Sky Area Multi-Object Fiber Spectroscopic Telescope survey \citep[LAMOST,][]{Deng2012RAA....12..735D}. Due to the fact that periods of RR Lyrae stars are quite short, and the ``first apparition'' only enjoys about $5\%$ of the whole period \citep{Chadid2011Carnegie}, long-time exposures and co-addition of spectra smooths the relevant emission features. But low-resolution and single-epoch spectra provided by SDSS and LAMOST make this work feasible.

The common practice to hunt the ``first apparitions'' is to visually check the profiles of hydrogen Balmer lines for any sign of emission for a small sample. As for large survey data, this becomes an enormous amount of work. So we built up a set of pipeline, using hand-crafted one-dimensional pattern recognition method. We presuppose the pattern of the target feature as a Gaussian-like emission profile and a broad Gaussian-like absorption profile when both the two signals are more significant than $2\sigma$ compared to the mean level. We use the minimum value in the selected windows to locate Balmer absorption profile. The ``hunting'' results are adopted which at least show clear patterns of the ``first apparition'' in $H\alpha$ and $H\beta$ at the same time and contain at least two observational points on the profiles of emission. We apply this pipeline to the spectra of RR Lyrae stars from SDSS and LAMOST, with visual checks to ensure completeness.

The emission and absorption lines are fitted based on the $scale$ $width$ $versus$ $shape$ method for the Balmer lines \citep{Sersic1968adga.book.....S,Clewley2002MNRAS.337...87C}. We adopted two $S\acute{e}rsic$ profiles \citep{Xue2008,Yang2014} as:
\begin{eqnarray}\label{eq:Sersicprofile}
y =m - a e^{-(\frac{\left|{\lambda}-{\lambda_0}\right|}{b})^c}
\end{eqnarray}
to fit the profile and make measurements of flux, intensity and full width at half maximum (FWHM) for the two components. We generate uncertainties by error propagation for covariance matrix and Monte Carlo method \citep{Andrae2010arXiv1009.2755A}, with limits based on observational resolutions.

\section{Results of Searching}\label{section:results}

\begin{figure*}[tb]
\begin{center}
\begin{tabular}{c}
\includegraphics*[width=70mm,height=4.65cm]{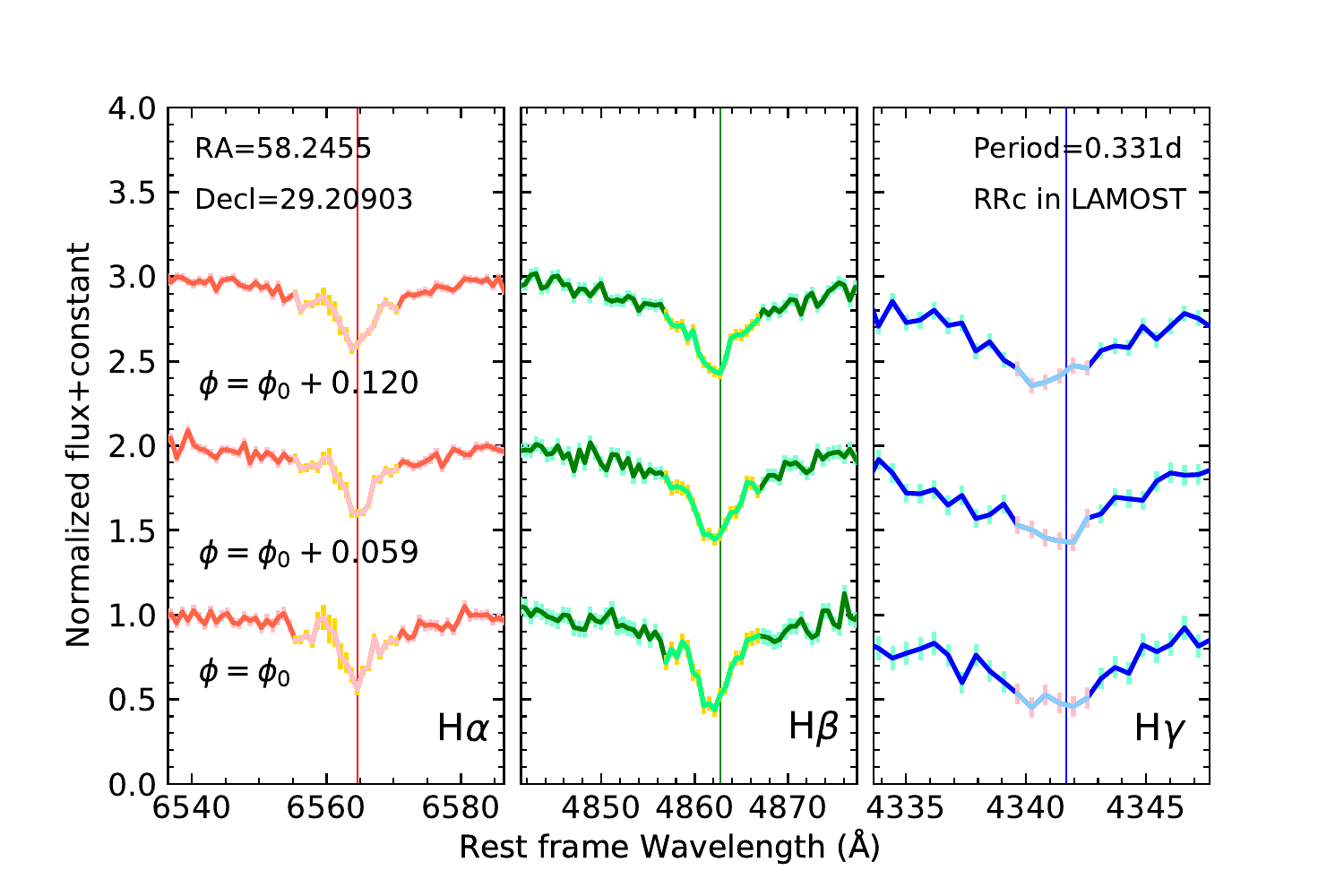}
\end{tabular}
\caption{\label{fig:opacityproblem2}
An example of the evolution of the H$\alpha$, H$\beta$, and H$\gamma$ line profiles. The frame of reference for the wavelength axis is the stellar rest frame. The vertical blue lines indicate the H$\alpha$, H$\beta$, and H$\gamma$ line laboratory wavelength.}\end{center}
\end{figure*}

\begin{figure*}[tb]
\begin{center}
\begin{tabular}{cc}
\includegraphics*[width=70mm,height=4.65cm]{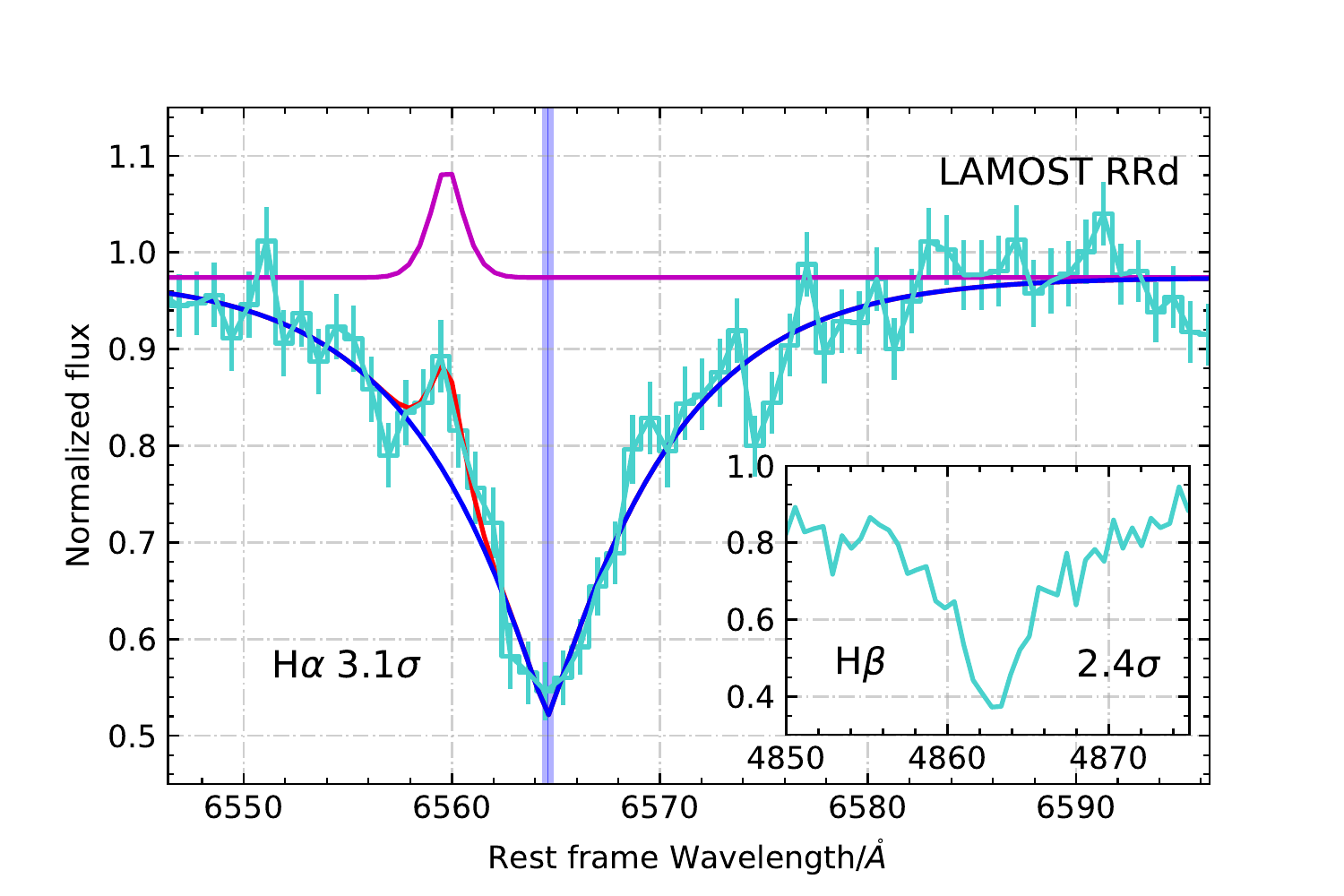}&
\includegraphics*[width=70mm,height=4.65cm]{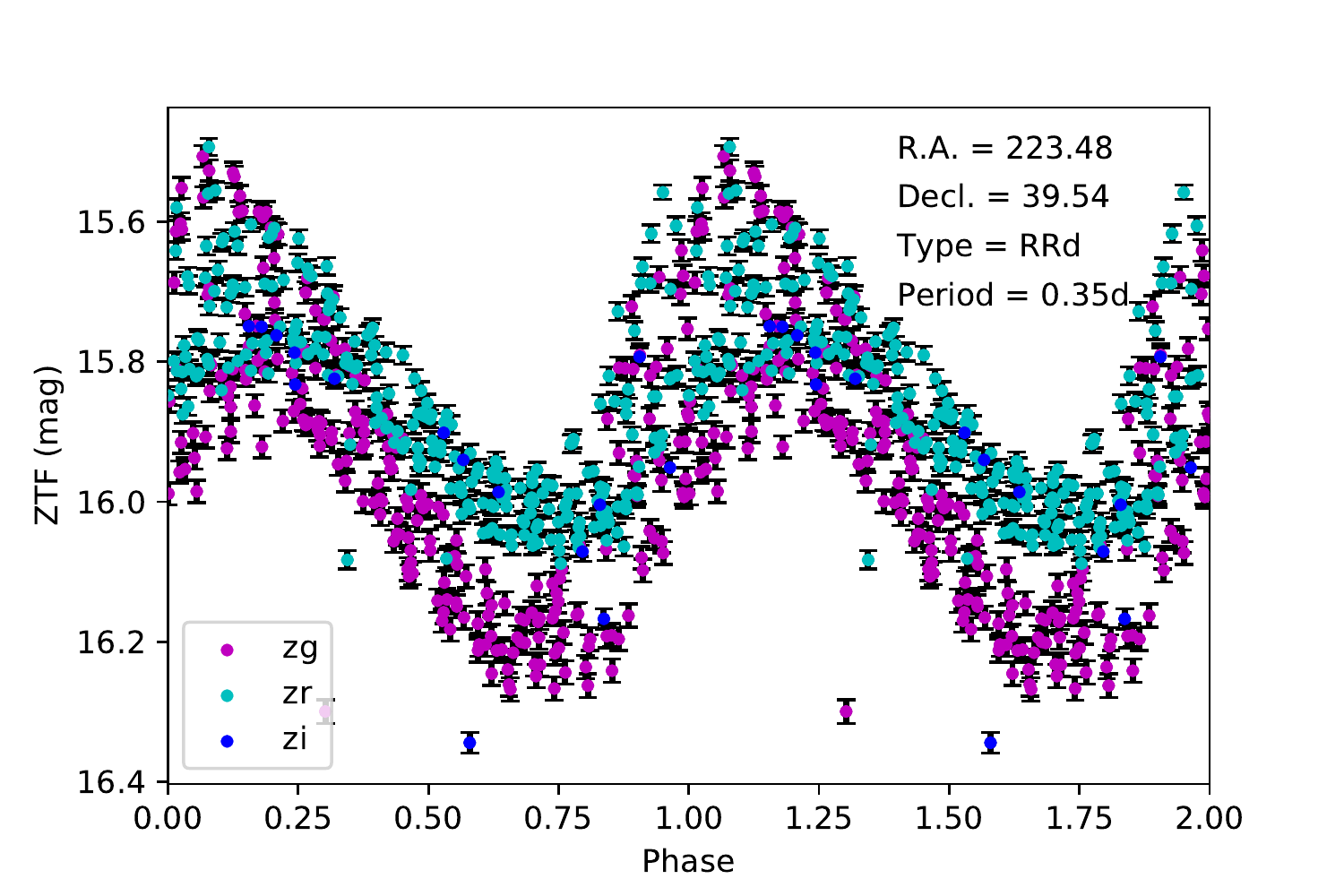}
\end{tabular}
\caption{\label{fig:opacityproblem3} 
Left (a): Fitting result of the ``first apparition'' of the RRd star that may suffer from Blazhko effect. 
The reference frame for the wavelength axis is the stellar rest frame. The H$\alpha$ emission lines are displayed as the pink profiles. The Vertical blue lines indicate the H$\alpha$ line laboratory wavelength. The H$\beta$ emission lines are displayed in the subplots. The significances of the emissions indicate the ratio between the flux of signal and noise.
Right (b): The light curve from ZTF of the RRd star that may suffer from Blazhko effect.} 
\end{center}
\end{figure*}

As one of the main results of the searching program, we find ten RRc stars in SDSS, ten RRc stars, and three RRd stars in LAMOST, which all show intense shifted emission components on the blue wing of the Balmer absorption lines.  
The flux is normalized by the fitting result of the continuum. 
Figure.~\ref{fig:opacityproblem2} presents an example of the evolution of line profiles. When the ``first apparition'' appears, the emission line appears at H$\alpha$, H$\beta$, and H$\gamma$ simultaneously.
Figure.~\ref{fig:opacityproblem3}(a) shows a fitting example of one RRd star in LAMOST with hydrogen emission.
Figure.~\ref{fig:opacityproblem3}(b) shows its light curve, provided by the Zwicky Transient Facility (ZTF). It may indicate that this RRd star suffers from long-term modulations (Blazhko effect).

\section{Discussion}\label{section:discussion}

Emission features shown in the spectra of RR Lyrae stars can be considered as shocks propagating through their pulsating atmosphere. The ``first apparition'' can be generated by a shock forming below the photosphere near the time of the minimum radius. The shock is accelerating outward. The atoms de-excite after being excited by the shock in the radiative wake. Energy from the ramp pressure at the shock front is high enough to excite neutral hydrogen from the second quantum state upwards. 
Studies \citep{Schwarzschild1952} show that hypersonic velocity shock is needed to make the emission observable. So we take our discoveries as evidence of the existence of hypersonic shock waves in non-fundamental mode RR Lyrae stars.

We measure the radial velocity of the emission line in the stellar rest frame. The equation is as:
\begin{eqnarray}\label{eq:Vshock}
V_{\rm e1,\alpha} = c\frac{(\lambda_{\rm e1,\alpha}-\lambda_{\rm ab})}{\lambda_{0}}
\end{eqnarray}
where $\lambda_{\rm e1,\alpha}$ is the wavelength corresponding to the central wavelength of the emission line. $\lambda_{\rm ab}$ indicates the central wavelength of the absorption component, while $\lambda_{0}$ is the laboratory wavelength.

An estimation of the temperature of the shock front for a strong shock can also be given by adiabatic Rankine-Hugoniot relationships \citep{Chadid2008A&A...491..537C} as:
\begin{eqnarray}\label{eq:RankineHugoniot}
V_{shock}=\sqrt{\frac{16}{3}\frac{R}{\mu}T_{shock}}
\end{eqnarray}
where $V_{shock}$ is in $km/s$, $\mu$ indicates the mean atomic weight, $R$ stands for the universal gas constant.

\begin{figure*}[tb]
\begin{center}
\begin{tabular}{c}
\includegraphics*[width=70mm,height=4.65cm]{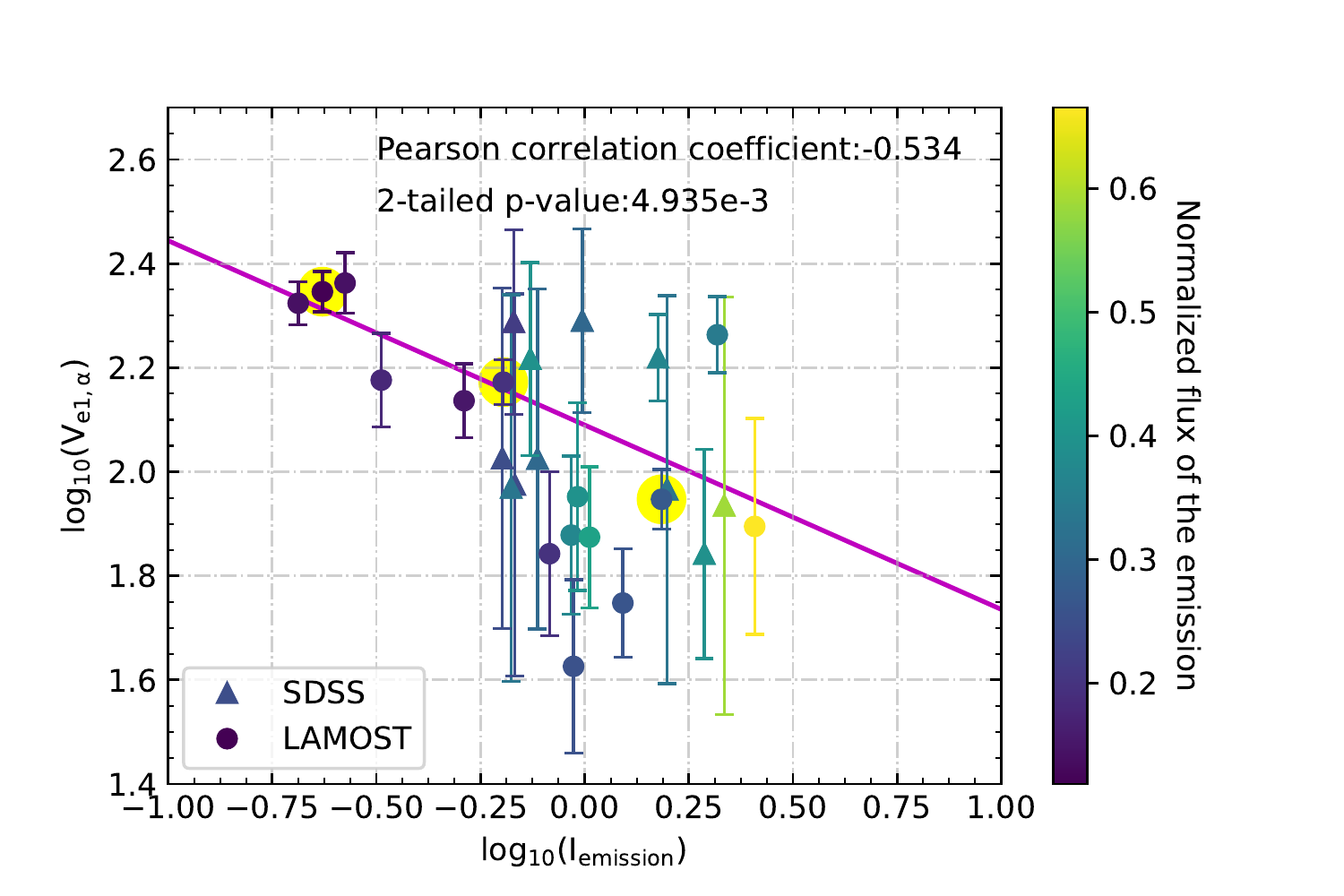}
\end{tabular}
\caption{\label{fig:opacityproblem4}
Relationship between the intensity and the radial velocity of the emission signal of the ``first apparitions''. The variation of the color shows different normalized flux of the emission part. Points highlighted with yellow edge are RRd stars, while others are RRc stars. Pink solid line indicates linear fitting results. Triangular points represent sample from SDSS, while circular points denote sample from LAMOST. The radial velocity of the emission is generated by the redshift measurements using $H\alpha$ lines.}
\end{center}
\end{figure*}

We investigate the relationship between the radial velocity and the intensity of the emission part of the ``first apparitions'', visualized in Figure.~\ref{fig:opacityproblem4}.  
It shows a clear anti-correlation, characterized by Pearson correlation coefficients of about -0.534. The variation of color shows that as the intensity of the emission becomes larger, the line width of the emission part gets broader, which indicates that the temperature of it gets higher.
When the Mach number increases, the shock wave is getting more and more drastic. But from our results, if the radial velocity of the emission can represent the velocity of the shock, we learn that the intensity of the emission of the ``first apparitions'' is actually decreasing as the velocity and the temperature of the shock front are increasing. A probable explanation of this contradiction between theory and observation is the increase of the opacity. In a helium ionization zone, like the envelope of RR Lyrae, the environment becomes more opaque as temperature rises, so the light is harder to be delivered. 
Since RRc and RRd stars are hotter population in the whole sample of RR Lyrae and the anti-correlation between emission feature intensity and radial velocity (Figure.~\ref{fig:opacityproblem4}), they are supposed to have less prominent shock signatures. It should be harder for them to deliver signatures of shock waves and it indicates that shock waves discovered in our sample are quite severe.

Theoretical analysis in Fokin 2012, private communications told us that the different types of shock waves identified in RRab can also be found in RRc stars. Main shocks generated by the $\kappa-\gamma-mechanism$ in RRc have an amplitude three to four times lower than those in RRab. The maximums of the flux of RRc is lower than RRab in our sample. The detailed comparisons between RRab and RRc sample with shock waves will be discussed in our upcoming paper. \cite{Gillet2013A&A...554A..46G} showed that statistically there is a striking amplitude jumping between stars with short-periods (RRc) and stars with long-periods (RRa) near the period $\sim 0.4$ day. According to our result, the contribution of opacity should also be taken into account.

The Blazhko effect is considered a common effect among RRab and RRc stars. It has been suggested that this effect is generated by several strong shocks occurring during each pulsation cycle \citep{Gillet2013A&A...554A..46G}. But it can't be confirmed that they share the same physical mechanism to suffer from this effect. It is mainly because of the lack of information on the observed shock wave signal in RRc. The large variety of resonant, nonresonant, and chaotic possible states should also be taken into considerations to interpret low-amplitude variations such as small modulations \citep{Moln2012ApJ...757L..13M,Moln2012AN....333..950M}. Apparently, our research will strongly contribute to the investigation between shock waves and long-term modulations in non-fundamental mode stars \citep{Duan2020}.

\vskip 0.5cm


\section{Conclusions}

We have detected the ``first apparitions'' in non-fundamental RR Lyrae stars for the first time, including ten RRc stars in SDSS sample, ten RRc stars, and three RRd stars in LAMOST sample. 
We find an anti-correlation relationship between the radial velocity and the intensity of the emission signal, which is contrary to the theoretical behavior of shock waves but can be explained as the result of the changing opacity in a helium ionization zone. 
Moreover, one of the RRd stars we have selected may suffer from long-term modulation. 

Nevertheless, due to the lack of observations of shock wave signals in non-fundamental mode RR Lyrae stars, the origin of the Blazhko effect in RRab and RRc stars has been suspected to be totally different. But since we have detected observational characteristics of hypersonic shock waves in non-fundamental mode stars, it may be possible for us to investigate the role of shock waves in long-term modulation of the first overtone and multi-mode RR Lyrae stars. As the cause of the formation of the Blazhko effect remains a mystery, steps toward the nature of shock waves may surprise us in the future. 

\vskip 1cm







\section*{Acknowledgements}
We appreciate the help from Dr. Hao-Tong Zhang, Dr. Zhong-Rui Bai and Dr. Jian-Jun Chen for accessing the single-epoch spectra from LAMOST. We acknowledge research support from the Cultivation Project for LAMOST Scientific Payoff and Research Achievement of CAMS-CAS. Li-Cai Deng acknowledges research support from the National Science Foundation of China through grants 11633005. Xiao-Dian Chen also acknowledges support from the National Natural Science Foundation of China through grant 11903045. Xiao-Wei Duan acknowledges support from the Peking University President Scholarship. Hua-Wei Zhang acknowledges research support from the National Natural Science Foundation of China (NSFC) under No. 11973001 and National Key R$\&$D Program of China No. 2019YFA0405504. 
 The Guoshoujing Telescope (the Large Sky Area Multi-Object Fiber Spectroscopic Telescope LAMOST) is a National Major Scientific Project built by the Chinese Academy of Sciences. Funding for the project has been provided by the National Development and Reform Commission. LAMOST is operated and managed by the National Astronomical Observatories, Chinese Academy of Sciences.

\vspace{5mm} 



\end{document}